\documentstyle[epsf,12pt]{article}
\begin{document}
\baselineskip=1.2\baselineskip
\title{Low-energy Antiproton Interaction with Helium}
\author{
W. R. Gibbs\\
New Mexico State University\\
Las Cruces NM, 88003}
\date{\today}
\maketitle

\abstract{
An ab initio potential for the interaction of the neutral helium atom 
with antiprotons and protons is calculated using the Born-Oppenheimer 
approximation. Using this potential, the annihilation cross section 
for antiprotons in the energy range 0.01 $\mu$eV to 1 eV is calculated.
}

\section{Introduction}

Recent work on the trapping of low-energy antiprotons \cite{holz} presented
the measurement of the
annihilation rate on helium atoms which remained in the partial vacuum. 
It was found that the rate increased sharply as the cooling took place and
then suddenly dropped to the background level.  There
has been some speculation concerning the cause of this sudden decrease in
the rate.  One possibility could be that there is a barrier of a few
millivolts which would only be important when the energy of the antiprotons
reaches this level.

While these observations formed the initial motivation for the work presented
here there are additional reasons why the knowledge of this potential is
interesting.

First, the experiment mentioned above points out the possibility of carrying
out experiments in which this annihilation cross section could be measured and,
in a more advanced version, the x-ray transitions might be measured as well.

Second, there are more general reasons for knowing this potential.  For example
in the particle remnants of the big bang there could well be some antiprotons 
left in space. To the extent that these antiprotons remained free they would
have thermalized with the ambient background of photons and so would have an
average energy corresponding to a temperature of around 2.7 K.  Their
interactions would be with naturally occurring hydrogen and helium.

As a third reason,  it is conceivable that there could exist pockets of
potential or perhaps changes in slope in the antiproton-atom potential which
could allow metastable states to be formed.  These might be associated with
configurations of the electrons which are strongly deformed from a spherical
distribution. Such pockets of  potential are perhaps more probable in larger
aggregates of matter but this simple case of helium provides an entry into the
subject. 

There exist some calculations \cite{voronin,hughes} of a nature related to this
reaction but the author knows of no calculation of this precise process.

The calculation of the annihilation cross section as a function of energy
requires the knowledge of the electrostatic potential between the $\bar{p}$ and
the  neutral helium atom, the annihilation potential of the nucleus and the
solution of the Schr\"odinger equation with these two potentials.  

It is natural to approach the problem of antiproton-helium  electrostatic
interaction, considering at the same time the proton-helium potential. In this
way a check on the calculation is provided.  The calculation of the
proton-helium ab initio (variational) potential  by Kolos and 
Peek\cite{kolos2} is considered to be the standard reference. 

While one might at first think that the calculations are almost the same (just
the charge of the external particle is changed), because of the art of choosing
the variational wave function they can be very different.  For instance, Kolos
and Peek (following earlier work by Kolos\cite{kolos1} and 
Wolniewicz\cite{wol}) chose a trial electron wave function which was elliptical
in form, thus naturally tending to surround the two positive charges.  This
function is quite appropriate for the two nuclei separated by small or
intermediate distances.  For large distances, where the electrons are nearly
spherically distributed around the helium nucleus, this wave function can be
expected to less efficient.  For the case of the antiproton this form is
perhaps less appropriate.

Since it is desirable to do both calculations on the same footing, a single
trial function has been use for both cases.  It is based on an expansion in
spherical coordinates about the helium nucleus.  This function is very
efficient for small or large distances between the two nuclei. It has  more
difficulty representing the wave function well at intermediate distances where
it may require a large number of terms.

In the following section the expressions for the calculation of the variational
ratio are derived.  In section \ref{formal} the results of the potentials are
given and in section \ref{ann} the annihilation cross section is computed.

\section{Calculation of the potentials\label{formal}}

The trial wave function chosen for this work is based on a spherical
expansion about the helium nucleus in terms of spherical harmonics and
orthogonal polynomials in the (scaled) radial distance of the electrons from
the center.  It is taken of the same form for the proton and antiproton
problem so that a direct comparison can be made.  This partial wave expansion 
goes over into the polarizability expansion naturally.  

\subsection{Variational Integrals}

A completely general form of the wave function  for two electrons can be
expressed as

\begin{equation} 
\psi({\bf r}_1,{\bf r}_2)=\sum \psi_{\ell_1,m_1,\ell_2,m_2,n_1,n_2}
Y_{\ell_1}^{m_1}(\hat{{\bf r}_1})Y_{\ell_2}^{m_2*}(\hat{{\bf r}_2})
L_{n_1}(y_1)L_{n_2}(y_2)e^{-\frac{1}{2}(y_1+y_2)} 
\end{equation}
where $L_n(r)$ is the Laguerre polynomial of order 2 (usual notation 
$L_n^{(2)}(r)$) and $y_1=r_1/a$, $y_2=r_2/a$, , ``a'' being a variational
parameter which sets the scale of the system. The quantities 
$\psi_{\ell_1,m_1,\ell_2,m_2,n_1,n_2}$ are the components of the wave
function in this basis and embody (reduced by one index, see below) the other
variational parameters of the calculation. It has been assumed, as usual
\cite{friedrich}, that the singlet spin state dominates the lowest
energy configuration leaving the spatial wave function symmetric.

Since the two-electron wave function can depend only on the relative value
of the $\phi$ angles, because of the symmetry around the $^4$He-$\bar{p}$ 
axis, the most general form reduces to

\begin{equation} 
\psi({\bf r}_1,{\bf r}_2)=\sum \psi_{m,\ell_1,\ell_2,n_1,n_2}
Y_{\ell_1}^m(\hat{{\bf r}_1})Y_{\ell_2}^{m*}(\hat{{\bf r}_2})
L_{n_1}(y_1)L_{n_2}(y_2)e^{-\frac{1}{2}(y_1+y_2)}.
 \end{equation}

It is this expression which will be used in the following work.  The sums on
$\ell$ and $n$ are taken to $\ell_{max}$ and $n_{max}$. The condition that the
spatial electron wave function is symmetric requires that the $\psi$'s are
symmetric under the interchange $(\ell_1,n_1)\leftrightarrow (\ell_2,n_2)$.

The normalization of the wave function is given by

\begin{equation} 
\sum (n_1+1)(n_1+2)(n_2+1)(n_2+2)\psi_{m,\ell_1,\ell_2,n_1,n_2}^2.
\end{equation}

The hamiltonian for the problem is

\begin{equation} 
H=-\frac{\hbar^2}{2m}\nabla^2_1-\frac{\hbar^2}{2m}\nabla^2_2
-\frac{2e^2}{r_1}-\frac{2e^2}{r_2}+\frac{e^2}{|{\bf r}_1-{\bf r}_2|}
\pm\frac{e^2}{|{\bf r}_1-{\bf R}|}\pm\frac{e^2}{|{\bf r}_2-{\bf R}|}, 
\end{equation}
where ${\bf R}$ is the vector separating the antiproton (proton) from
the helium nucleus.

In order to carry out the variational calculation the computation of the 
expectation value of the trial wave function of each of these terms is
needed.

The radial derivative part of the kinetic energy can be expressed as

\begin{equation} 
 \frac{-\hbar^2}{2ma^2}K(n_1,n'_1,n_2,n'_2)\delta_{\ell_1,\ell'_1}
\delta_{\ell_2,\ell'_2}\delta_{m,m'}   
\end{equation}
where $K$ is an integer.

Twice the contribution to the kinetic energy for one of the electrons is
given by
\begin{equation} 
KK(n',n)=\delta_{n,n'}\frac{(n+1)(n+2)}{2}-2na_{n',n}+2(n+2)a_{n',n-1}
-2(n+1)b_{n',n} 
\end{equation}
so that the contribution to both will be
$$ K(n_1,n'_1,n_2,n'_2)=\frac{1}{2}\left[KK(n'_1,n_1)(n_2+1)(n_2+2)\delta_{n'_2,n_2}
\right.$$
\begin{equation} 
\left.+KK(n'_2,n_2)(n_1+1)(n_1+2)\delta_{n'_1,n_1}\right] ,
\end{equation}
with
\begin{equation} 
a_{n_1,n_2}\equiv \int_0^{\infty} e^{-y}L_{n_1}(y)L_{n_2}(y)dy 
\end{equation}
\begin{equation} 
a_{n,n}=\frac{(n+1)(n+2)(2n+3)}{6};\ \ a_{n,n+m}=a_{n,n}+
\frac{m(n+1)(n+2}{2} 
\end{equation}

\begin{equation} 
b_{n_1,n_2}\equiv \int_0^{\infty} ye^{-y}L_{n_1}(y)L_{n_2}(y)dy 
 =\frac{(n+1)(n+2)}{2};\ \ \ \ n={\rm min\ of\ } (n_1,n_2) 
\end{equation}

The angular momentum part contributes

$$ K_L(m,m',\ell_1,\ell'_1,\ell_2,\ell'_2,n_1,n'_1,n_2,n'_2)= 
\delta_{m,m'}\delta_{\ell_1,\ell'_1}\delta_{\ell_2,\ell'_2} $$

\begin{equation}  
\times \left[
\ell_1(\ell_1+1)(n_2+1)(n_2+2)a_{n_1,n'_1}\delta_{n'_2,n_2}+
\ell_2(\ell_2+1)(n_1+1)(n_1+2)a_{n_2,n'_2}\delta_{n'_1,n_1} \right].
\end{equation}

The contribution to the potential energy of the electron-He interaction
is given by

\begin{equation} 
\frac{
2e^2\delta_{m,m'}\delta_{\ell_1,\ell'_1}\delta_{\ell_2,\ell'_2}}{a}
\left[(n_1+1)(n_1+2)b_{n_2,n'_2}\delta_{n_1,n'_1}+
(n_2+1)(n_2+2)b_{n_1,n'_1}\delta_{n_2,n'_2}\right] .
\end{equation}

The expectation value of the electron-electron interaction involves
calculating

\begin{equation} 
\frac{e^2}{a}\int d{\bf y}_1d{\bf y}_2 Y_{\ell'_1}^{*m'}(\hat{{\bf y}}_1)
Y_{\ell'_2}^{m'}(\hat{{\bf y}}_2)
L_{n'_1}(y_1)L_{n'_2}(y_2)\frac{e^{-(y_1+y_2)}}{|{\bf y}_1-{\bf y}_2|}
Y_{\ell_1}^{m}(\hat{{\bf y}}_1)Y_{\ell_1}^{*m}(\hat{{\bf y}}_2)
L_{n'_1}(y_1)L_{n'_1}(y_1).
\end{equation}

\begin{equation} 
\frac{e^2}{a}\sum \sqrt{\frac{(2\ell_1+1)(2\ell_2+1)}
{(2\ell'_1+1)(2\ell'_2+1)}}C_{\ell_1,L,\ell'_1}^{0,0,0}
C_{\ell_2,L,\ell'_2}^{0,0,0}C_{\ell_1,L,\ell'_1}^{m,M,m'}
C_{\ell_2,L,\ell'_2}^{m,M,m'} I(n_1,n'_1,n_2,n'_2,L),
 \end{equation}

where

\begin{equation} 
I(n_1,n'_1,n_2,n'_2,L)\equiv
\int y_1^2 dy_1 y_2^2 dy_2 e^{-(y_1+y_2)}L_{n'_1}(y_1)L_{n'_2}(y_2)
\frac{y_<^L}{y_>^{L+1}}L_{n_1}(y_1)L_{n_2}(y_2) 
\end{equation}

The coefficient for the expansion of a product of Laguerre polynomials

\begin{equation} L_{n_1}(y)L_{n_2}(y)=\sum A_{n_1,n_2,n_3}L_{n_3}(y) \end{equation}
can be found by using the explicit expression of the Laguerre polynomials

\begin{equation} 
L_n(y)=\sum_{m=0}^n (-1)^m\left(\begin{array}{c} n+2\\n-m\end{array}\right)
\frac{y^m}{m!}= \sum_{m=0}^n d_{n,m}y^m
\end{equation}

$$(n_3+1)(n_3+2)A_{n_1,n_2,n_3} =(n_1+2)!(n_2+2)!(n_3+2)! $$

\begin{equation} 
\times\sum_{m_1,\ m_2,\ m_3=0}^{n_1,\ n_2,\ n_3}
\frac{(-1)^{m_1+m_2+m_3}(m_1+m_2+m_3+2)!}{
(n_1-m_1)!(m_1+2)!m_1!(n_2-m_2)!(m_2+2)!m_2!(n_3-m_3)!(m_3+2)!m_3!} 
\end{equation}

With the definition

\begin{equation} 
I_{k,n}=\int_0^{\infty}y^ke^{-y}dy\int_0^yx^ne^{-x}dx 
\end{equation}

\begin{equation} 
I_{k,n}=\left\{\begin{array}{l} 
n!\sum_{m=n+1}^{\infty}\frac{(k+m)!}{m!2^{m+k+1}}
\ \ \ \ \ \ \ k\le 0 \\
 \\
n!\left[ k!-\sum_{m=0}^n\frac{(k+m)!}{m!2^{m+k+1}}\right]
\ \ \ k\ge 0 \end{array}\right\} 
\end{equation}
we have

$$I(n_1,n'_1,n_2,n'_2,L)=$$
\begin{equation}
\sum A_{n_1',n_1,j_1}A_{n_2',n_2,j_2} d_{j_1,k_1}
d_{j_2,k_2} \left[I_{k_1+L-1,k_2-L+1}+I_{k_2+L-1,k_1-L+1}\right]. 
\end{equation}

To evaluate the interaction energy with the antiproton (proton) we need

\begin{equation} 
\int_0^{\infty} \frac{y^ne^{-y}}{|{\bf y} -{\bf Y} |}dy=
\sum P_L(\cos\theta)\left[\int_0^Y\frac{y^{n+L}}{Y^{L+1}}e^{-y}dy
+\int_Y^{\infty}y^{n-L-1}Y^Le^{-y}dy \right]
\end{equation}

\begin{equation}
 =Y^n\sum P_L(\cos\theta)\left[\int_0^1t^{n+L}e^{-Yt}dt
+\int_1^{\infty}t^{n-L-1}Y^Le^{-Yt}dt \right]
\end{equation}
where Y is the distance from the He nucleus to the antiproton in units of
$a$ along the z-axis.

If we define

\begin{equation} 
\alpha_m(Y)\equiv \int_1^{\infty} t^me^{-Yt}dt;\ \ \ m\ge 0, 
\end{equation}
then

\begin{equation} 
\alpha_0(Y)=\frac{e^{-Y}}{Y};\ \ \ Y\alpha_m(Y)=e^{-Y}+m\alpha_{m-1}(Y)
. \end{equation}

Note

\begin{equation} 
\int_0^1t^me^{-Yt}dt=\frac{m!}{Y^{m+1}}-\alpha_m(Y)\equiv \gamma_m(Y)
\end{equation}

If $m<0$ then the exponential integral is needed

\begin{equation} 
E_m(Y)=\int_1^{\infty} t^{-m}e^{-Yt}dt;\ \ \ E_{m+1}=\frac{[e^{-Y}-YE_m(Y)
]}{m} 
\end{equation}

$E_1(Y)$ must be calculated to high accuracy.

With the definition
\begin{equation} 
\beta_m(Y)\equiv \left\{\begin{array}{l} \alpha_m(Y)\ \ \ m\ge 0\\
E_{-m}(Y)\ \ \ m<0\end{array}\right\} 
\end{equation}
we can write

\begin{equation} 
\int_0^{\infty}\frac{y^ne^{-y}}{|{\bf y}-{\bf y}|}=\sum P_L(\cos\theta)
Y^n[\gamma_{n+L}(Y)+\beta_{n-L-1}(Y)] 
\end{equation}
and

$$\int d{\bf y} \frac{Y_{\ell '}^{*m'}(y)Y_{\ell}^m(y)L_{n'}(y)L_{n}(y)}
{|{\bf y} -{\bf y}|}=$$

\begin{equation}
 =\delta_{mm'}\sum A_{n,n',\bar{n}}d_{\bar{n},m}C_{\ell,L,\ell '}^{0,0,0}
C_{\ell,L,\ell'}^{m,0,m}\sqrt{\frac{2\ell +1}{2\ell '+1}} 
\left[\gamma_{m+2+l}(Y)+\beta_{m-L+1}(Y)\right] 
\end{equation}

\begin{figure}[htb]
\epsfysize=100mm
\epsffile{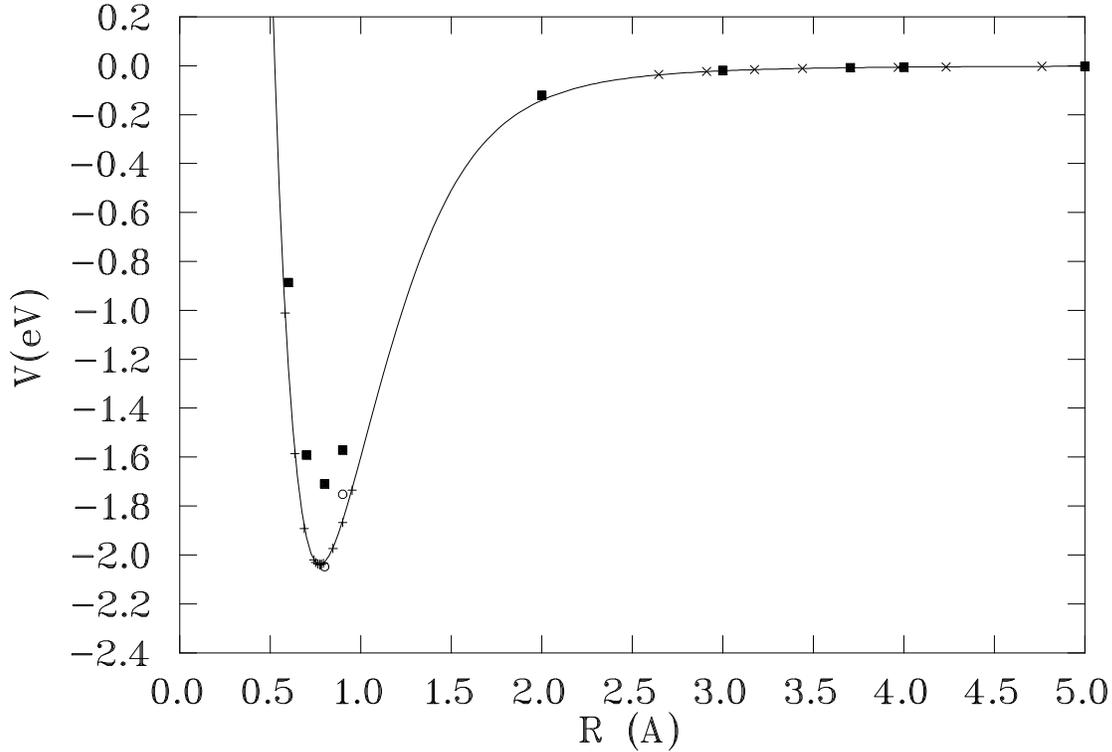}

\caption{Proton helium potential. The solid line is from
Ref. \protect{\cite{bosanac}}. The solid boxes are from the (4,4)
calculation, the open circles from the Pad\'e extension and the crosses
and x symbols are from \protect{\cite{kolos1,kolos2}}.
}\label{ppot}\end{figure}

\begin{figure}[htb]
\epsfysize=100mm
\epsffile{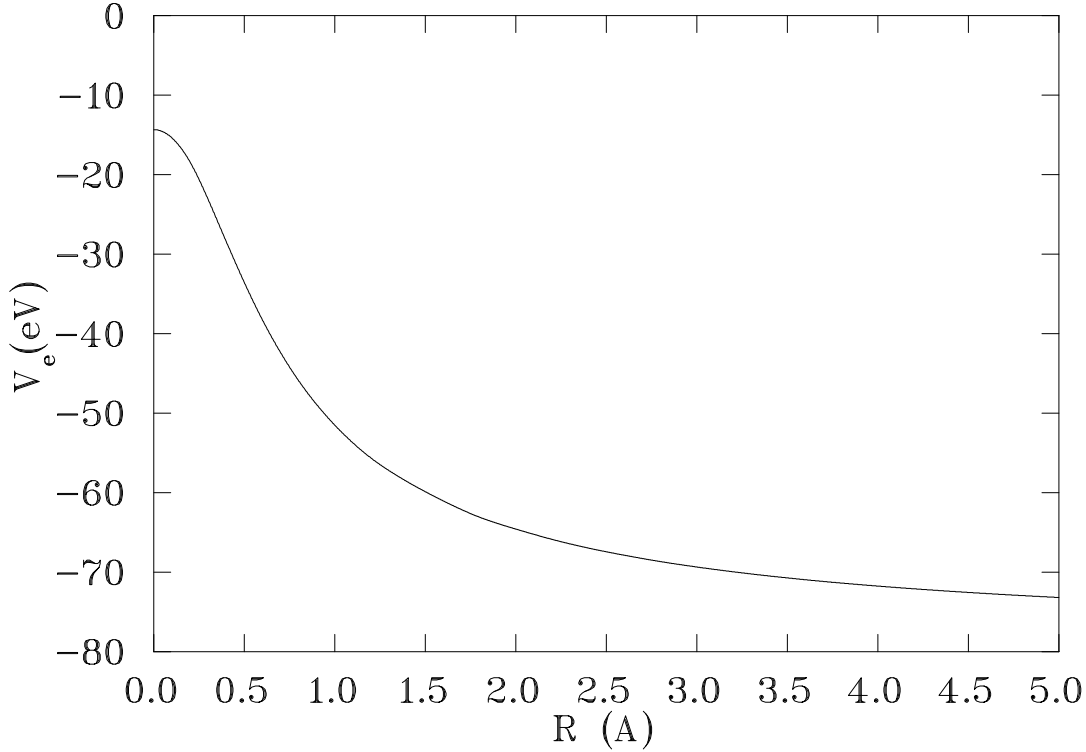}

\caption {Potential energy of the two-electron system.}\label{heapee}
\end{figure}

\subsection{Results for the proton potential\label{pot}}

The results for the proton potential are shown in Fig. \ref{ppot}. Calculations
were made at various radii for the pair $(\ell_{max},n_{max})=(2,2),\ 
(3,3)$ and $(4,4)$.  The (4,4) results (solid squares) are shown in the
figure.  Also shown are the results of ref. \cite{kolos1,kolos2} and the 
parameterization of their potential by Bosanac and Knesaurek \cite{bosanac}.
The (4,4) results give an adequate representation of the previous results
except in the minimum where there is a significant cancellation 
between the electronic potential energy and the direct proton-nucleus 
interaction.  

A [1,1] Pad\'e approximate was used with the three determinations mentioned
above to estimate the result of the limit of the sequence. This result
is shown as the open circles in Fig \ref{ppot}. While this last extrapolation
is significant the result gives a satisfactory agreement with the previous
work.

A specific comparison was made with ref. \cite{kolos1,kolos2} at R=3.704 
\AA . They found --8.32 meV while the present work gives --8.10 meV  for the 
(4,4) search.

\begin{figure}[htb]
\epsfysize=100mm
\epsffile{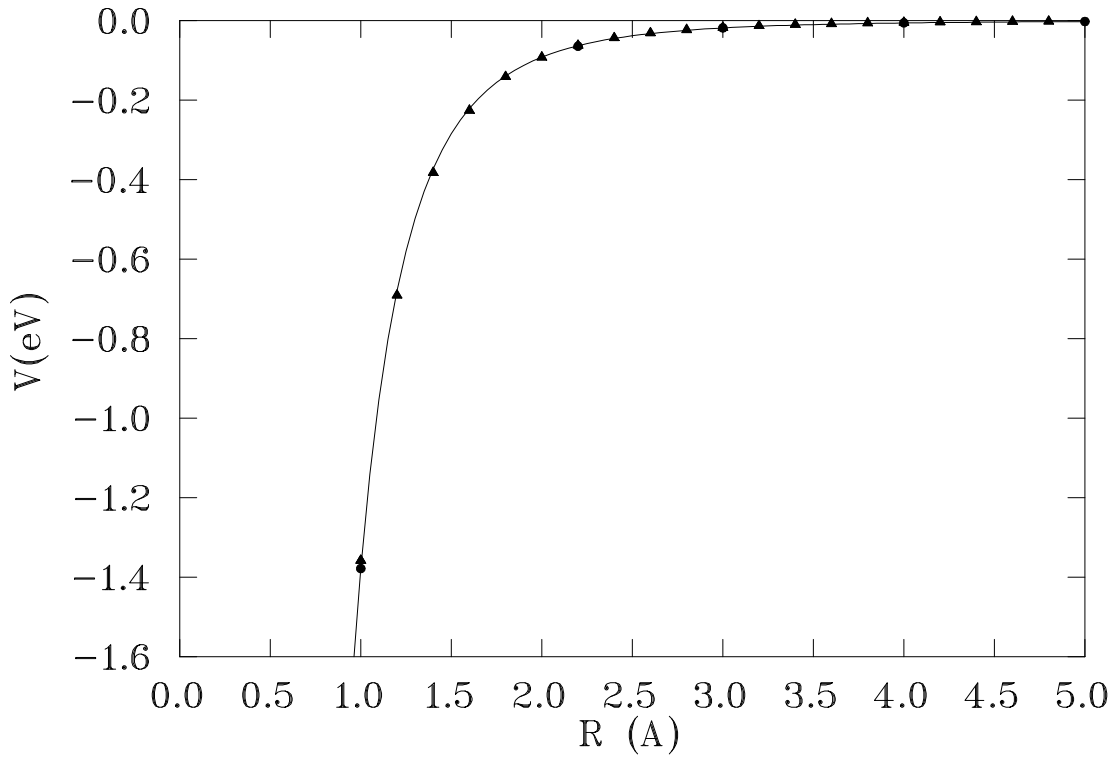}
\caption{Antiproton He potential. The triangles are the result of the
(3,3) calculation and the solid circles are from the (4,4) case.
 The solid line is calculated from Eq. \protect{\ref{pfit}}.}\label{appot1}
\end{figure}

\begin{figure}[htb]
\epsfysize=100mm
\epsffile{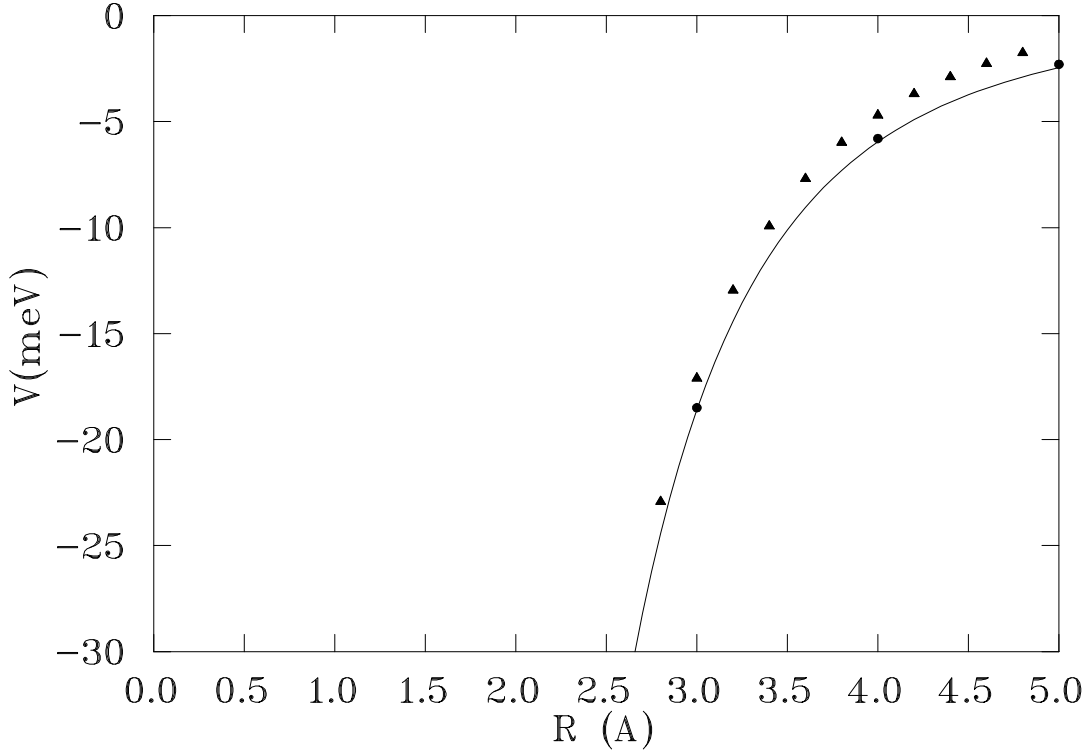}
\caption{Antiproton He potential. The meaning of the symbols is the same
as in Fig. \protect{\ref{appot1}}
}\label{appot2}
\end{figure}

\begin{figure}[htb]
\epsfysize=100mm
\epsffile{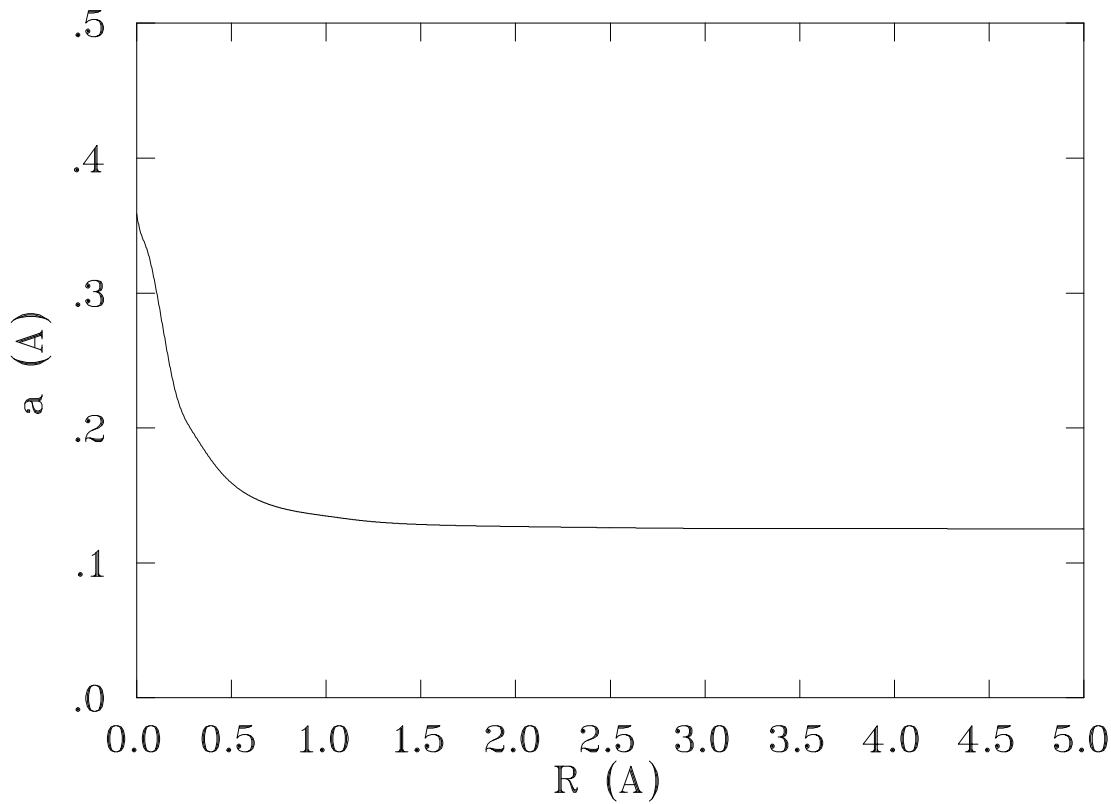}
\caption{Variational scale parameter ``a''.}\label{scale}
\end{figure}

\begin{figure}[htb]
\epsfysize=100mm
\epsffile{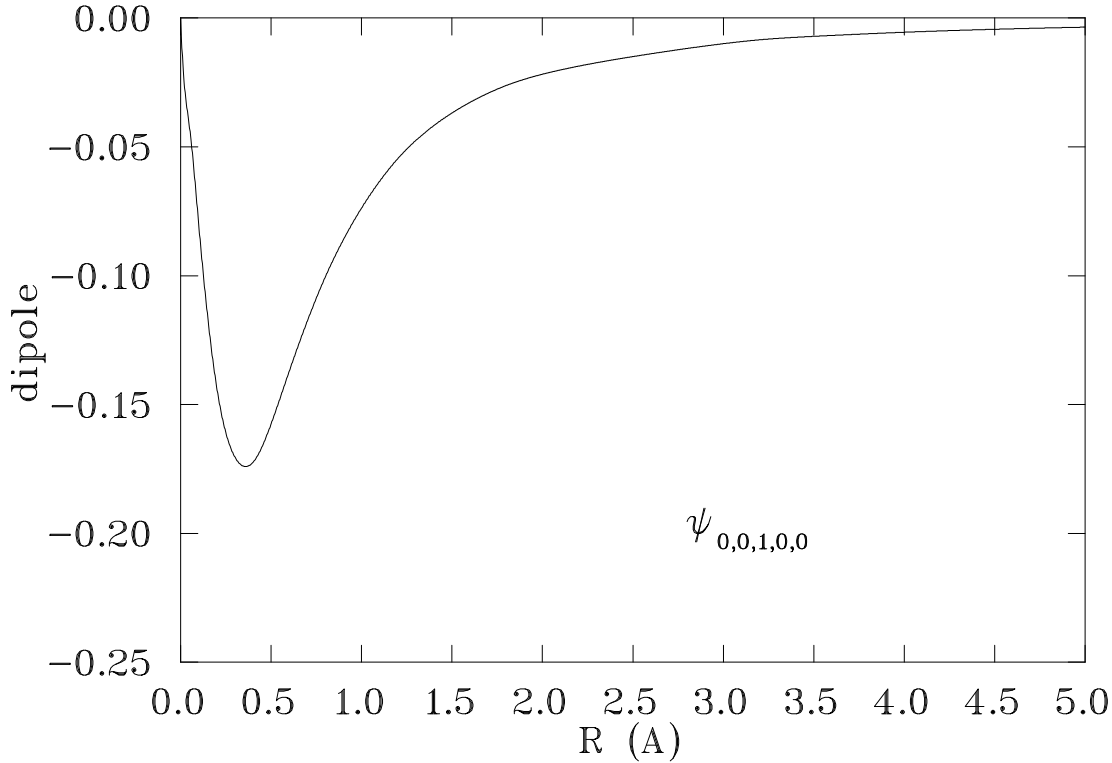}
\caption{Dipole coefficient in the wave function.}\label{dipole}
\end{figure}

\subsection{Results for the antiproton potential}

The potential energy for the electrons alone is shown in Fig. \ref{heapee}.
It is seen to vary between the limits $E_0$ at $R=0$ (approximately equal
to the binding energy of $H^-$) to $E_1$ (the binding energy of isolated
helium) at large distances.

Writing the electron potential as a ratio of two fourth-order polynomials,
 including the direct He-$\bar{p}$ electrostatic potential,
 assuming that the potential varies as $1/R^4$ at large values of $R$ and
subtracting the asymptotic value, $E_1$, 
the full antiproton-helium atom electrostatic potential can be expressed as

\begin{equation}
 V(R)=-\frac{\beta z(z^3+\frac{d}{E_0-E_1}z^4)}
{1+gz+\frac{\beta}{E_0-E_1}z^4}\label{pfit} 
\end{equation}
where $z\equiv 1/R$.

In the fit $ E_0=-14.3477$  eV and $E_1=-78.9847$ eV were used. The constant
``d'' ($=e^2$) has the value 28.798 eV $\AA^{-1}$ .
By searching on  parameters in
this form against the (4,4) results, a best fit was
found for $g=0.6474$ \AA\  and $\beta =1595 $ meV-\AA$^4$. The last
value can be compared with 1557 meV-\AA$^4$ from Ref \cite{davison} and
1678 meV-\AA$^4$ from Ref \cite{singh}.

The antiproton potential is shown in Figs. \ref{appot1} and \ref{appot2}.
It is seen that there is no apparent change in slope which might aid in the
formation of a metastable state.  Since any such effect would be expected to be
small (perhaps in the few meV range) it would have to be active at moderately
large distances to be visable, say beyond 3 \AA . We can perhaps understand 
why it is likely that there is 
no such effect by looking at the structure of the electron wave function.

The variation in the 
 scale parameter ``a'' with R is shown if Fig. \ref{scale}.  It is
seen that the system undergoes a rapid growth for a baryonic separation inside
of 1 \AA .
Figure \ref{dipole} shows the first dipole component of the wave function which
provides a measure of the deformation of the electron cloud from spherical
symmetry.
It also shows little effect outside of 1 \AA .  Thus any possible inflection in
the potential curve would occur inside of 1 \AA\ when the potential is 
completely dominated by
the simple coulomb attraction of the He-$\bar{p}$ system.

\section{Annihilation cross section\label{ann}}

The energy of the antiproton at the nuclear surface will be given by the
incident energy plus the gain due to the acceleration in the coulomb
potential.  Since (at 1 fm) the coulomb energy is 2.88 MeV, in the range
of incident energies considered here the annihilation takes place at
constant energy to a good approximation.  The nuclear potential was
taken to be the product of a purely absorbed single nucleon strength with
a Woods-Saxon density with radius 2 fm and diffuseness 0.5 fm for
4 nucleons.

The system was treated from the point of view of a nuclear optical 
model, i.e. no consideration was
given to the possibility of the knock-out of electrons or the electromagnetic
transition into atomic bound states.  In general such corrections might be
expected to increase slightly the annihilation cross section.

The problem could be treated in a perturbation approach. In that case
one would solve the problem of scattering from the purely electrostatic
potential generated in the previous section and calculate the annihilation
rate from the expectation value of the potential just introduced.  Since
the nuclear potential is very short ranged, the amplitude will be
proportional to the square of the wave function at the origin. It
was found in calculating this quantity that
it had very nearly the same value as the pure coulomb case

\begin{equation}
 C^2_0(\eta )=2\pi \eta (e^{2\pi\eta} -1)^{-1}\approx -2\pi \eta 
;\ \ \ \eta =-\alpha c/v 
\end{equation}
 for energies above 10 $\mu V$. From these considerations it is seen that
the cross section should vary as 1/$v^2$.  Since the potential is strongly
absorbing the wave function will be modified significantly in the 
region of 
the nuclear potential so that it is better to solve the full equation
as now described.

Since there exists no known analytic form for the solution of the Schr\"odinger
equation in these potentials at short range or in the
region of the ``surface'' of the atom, the wave function was calculated
over the entire region from 0 to 80 \AA . The step size was changed by an
order of magnitude 4 times with 20,000 points calculated in each of the
first four (overlapping) regions and 160,000 points in the region of
largest distance from the helium nucleus.

\begin{figure}[htb]
\epsfysize=100mm
\epsffile{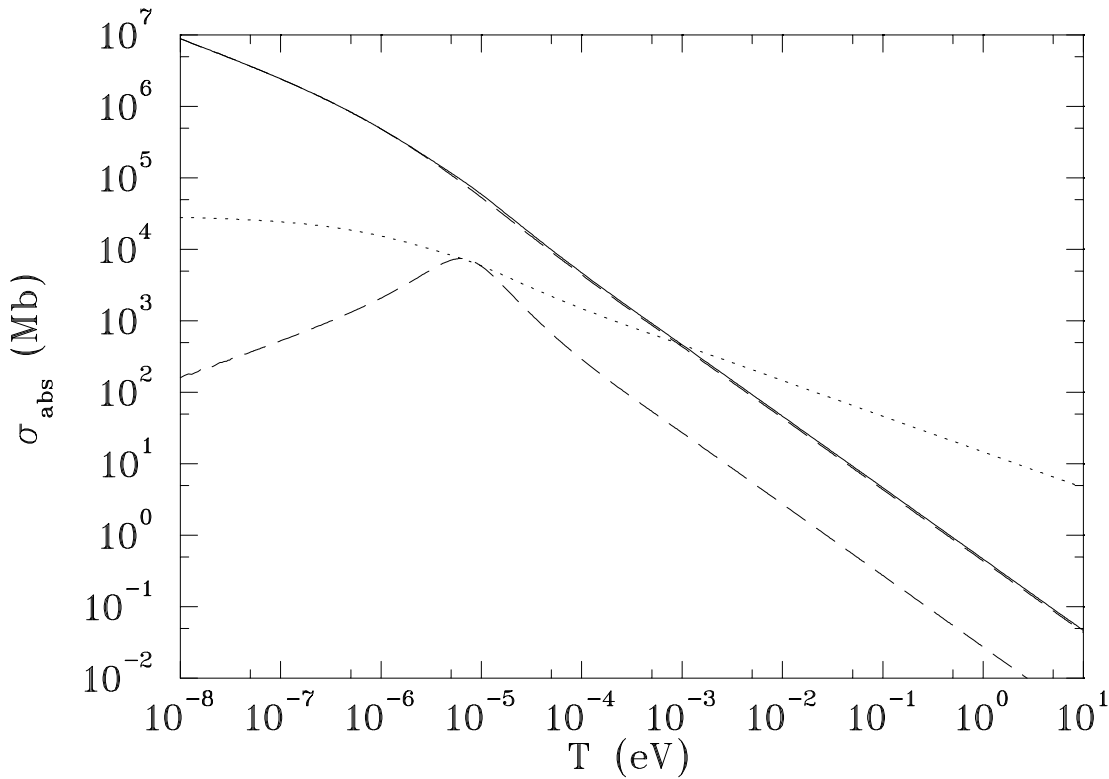}

\caption{Absorption cross section for antiprotons incident on neutral helium
atoms.  The solid curves gives the result due to the sum of all partial waves
while the separated dashed curve gives the p-wave contribution. The dashed
curve almost coincident with the solid curve represents the s-wave contribution
alone. The short dashed curve gives the product of the velocity and the cross
section normalized to the cross section at 1 meV.}  \label{sigma}\end{figure} 

The results for the annihilation cross section are presented in Fig. 
\ref{sigma}. It is seen that
there is a change in energy dependence around $10 \mu V$.  Above that energy
the cross section is proportional to $1/v^2$ as can be expected for
a coulomb system.  Below that energy the cross section varies as
$1/v$ as expected for a low-energy neutral system.

One can estimate this transition point using the asymptotic form
of the potential.  We can define a radius representing the ``surface''
of a potential as that value where the incident wave number is
significantly affected by the potential, say where the
potential becomes a fraction, f, of the incident kinetic energy. In the
present case the radius $R_0$ will be defined by
\begin{equation} 
E=\frac{k^2}{2m}=f\beta/R_0^4. 
\end{equation}
The long wave length limit is reached at $kR_0<1$.
Combining these conditions, it is seen that the transition should
take place around 2/f $\mu$ V or about 20 $\mu V$ for f=0.1.

The s-wave cross section above $~10^{-5}$ eV can be represented by the 
expression
\begin{equation} 
\sigma_{abs}=\frac{0.536\pi}{k^2} \label{totabs}
\end{equation}
Since this is greater than half of the unitarity limit the full
cross section cannot be much larger than this. While the overall accuracy of
the present calculation does not justify the three significant figures 
quoted in Eq. \ref{totabs}, the result is stable to this precision in 
this energy range.
Thus if the absolute cross section can be established at any one value
of the energy it 
will be known over a wide range. The measurement of the annihilation rate
on a small, known, amount of helium introduced into a trap could thus 
provide a direct measurement of the temperature over a certain region
of energy.

\section{Conclusion}

The authors in Ref. \cite{holz} estimate that the pressure in the container 
was of the order of $10^{-11}$ Torr.  If we assume that the helium and
antiprotons are cooled to 4.2 $^{\circ}$K then the annihilation rate
calculated using the cross section from Fig. \ref{sigma} is $\sim 1.2\times
10^{-3}$ 1/sec.  Since a maximum rate of $\sim 8\times 10^{-3}$ 1/sec was
observed, there is a slight discrepancy.  Possible explanations of this 
discrepancy include the existance of a higher pressure than estimated or 
the presence of heavier material.

It was seen that no barrier or inflection of the potential occurred.  However,
the
possible corrections beyond the Born-Oppenheimer approximation have not
yet been considered and could influence this conclusion.

The author gratefully acknowledges discussions with T. Goldman, M. Nieto
and M. H. Holzscheiter.

This work was supported by the U. S. Department of Energy.

\end{document}